\documentclass[12pt]{article}
\usepackage{psfrag}
\usepackage{graphicx}
\usepackage{graphics}
\usepackage{epsfig}
\newcommand{\As}{A\!\!\!/}
\newcommand{\ks}{k\!\!\!/}
\newcommand{\ps}{p\!\!\!/}

\date{}
\begin{document}
\baselineskip=18.6pt plus 0.2pt minus 0.1pt \makeatletter
%\@addtoreset{equation}{section} \renewcommand{\theequation}{\thesection.%
%\arabic{equation}} \makeatletter \@addtoreset{equation}{section}

%\begin{titlepage}

\title{\vspace{-3cm}
\hfill\parbox{4cm}{\normalsize \emph{}}\\
 \vspace{1cm}
{Laser-Assisted Semi Relativistic Excitation of Atomic Hydrogen by Electronic Impact.}}
 \vspace{2cm}

\author{S. Taj$^{1}$, B. Manaut$^{1}$\thanks{{\tt manaut@fstbm.ac.ma}}, M. El Idrissi$^{1}$ and L. Oufni $^{2}$\\
{\it {\small $^1$ Facult\'e Polydisciplinaire, Universit\'e Sultan Moulay Slimane,}}\\
{\it {\small Laboratoire Interdisciplinaire de Recherche en Sciences et Techniques (LIRST),}}\\
 {\it {\small Boite Postale 523, 23000 B\'eni Mellal, Morocco.}}\\
 {\it {\small $^{2}$ Universit\'e Sultan Moulay Slimane, Facult\'e des Sciences et Techniques,}}\\
{\it {\small  D\'epartement de Physique, LPMM-ERM, Boite Postale
523, 23000 B\'eni Mellal, Morocco.}}}
   \maketitle \setcounter{page}{1}
\begin{abstract}
The excitation of H ($1s-2s$) by electron impact in the presence and in the absence of the laser field is studied in the framework of the first Born approximation. The angular variation of the laser-assisted differential cross section (DCS) for atomic hydrogen by electronic impact is presented at various kinetic energies for the incident electron. The use of Darwin wave function as a semirelativistic state to represent the atomic hydrogen gives interesting results when the condition $z/c\ll1$ is fulfilled. A comparison with the non relativistic theory and experimental data gives good agreement. It was observed that beyond ($2700$ $eV$) which represents the limit between the two approaches, the non relativistic theory does not yield close agreement with our theory and that, over certain ranges of energy, it can be in error by several orders of magnitude. The sum rule given by Bunkin and Fedorov and by Kroll and Watson \cite{22} has been verified in both nonrelativistic and relativistic regimes.
\vspace{.04cm}\\
PACS number(s): 34.80.Dp, 12.20.Ds
\end{abstract}

\maketitle
%---------------------------------
\section{Introduction}
%----------------------------------
Relativistic laser-atom physics has become recently as a new research area and constitute a new field of systematic experimental and theoretical study. In particular, the laser-asssited electron-atom scattering becomes a rapidly growing subject. These processes are however of fundamental interest and paramount importance, for instance in the laser heating of plasmas and high-power gaz lasers. One of its most remarkable features is the possibility of exciting the target via the absorption of one or more photons. In recent years considerable attention has been given to these topics. Most of the works are theoretical and there is very little experimental work at present. Several experiments have been performed, in which the exchange of one or more photons between the electron-target and the laser field has been observed in laser-assisted elastic [1] and inelastic scattering [2-5]. In particular, the excitation processes have been largely investigated in the literature by several authors [6-9], mainly in the perturbative weak-field limit. The first theoretical studies on the inelastic scattering were inspired from the pioneering works [10-12], in which the interaction between the free electron and the field can be treated exactly by using the exact Volkov waves [13]. Another investigation have been done for the exact analytical relativistic excitation $1S_{1/2}\longrightarrow 1S_{1/2}$ of atomic hydrogen, by electron impact in the presence of a laser field, see [14].  For a summary of relativistic laser-atom collisions, see [15]. In the absence of the laser field, many authors have studied this process using numerical tools. Thus, Kisielius \textit{et al.} [16] employed, the R-matrix method with nonrelativistic and relativistic approximations for the hydrogen like $He^+$, $Fe^{25+}$ and $U^{91+}$ ions, where the case of transitions $1s\longrightarrow 2s$ and $1s\longrightarrow 2p$ as well as those between fine structure $n=2$ levels was considered. Andersen \textit{et al.} [17]  have applied the semirelativistic Breit Pauli R-matrix to calculate the electron-impact excitation of the $^2S_{1/2}$ $\longrightarrow$ $^2P^o_{1/2,3/2}$ resonance transitions in heavy alkali atoms. Payne \textit{et al.} [18]  have studied the electron-impact excitation of the $5s\longrightarrow 5p$ resonance transition in rubidium by using a semi-relativistic Breit Pauli R-matrix with pseudo-states (close-coupling) approach. S. Taj \textit{et al.} [19] have presented a theoretical semirelativistic model and have found that the semirelativistic Coulomb Born approximation (SRCBA) in a closed and exact form for the description of the ionization of atomic hydrogen by electron impact in the first Born approximation is valid for all geometries.
The aim of this contribution is to add some new physical insights and to  give the analytical formula for the excitation differential cross section for hydrogen from the 1s ground state to the 2s excited state in the absence and in the presence of the laser field. Before we present the results of our investigation, we first begin by sketching the main steps of our treatment. For pedagogical purposes, we begin by the most basic results of our work using atomic units (a.u) in which one has ($\hbar=m_e=e=1$), where $m_e$ is the electron  mass at rest, and which will be used throughout this work. We will also work with the metric tensor $g^{\mu\nu}= diag(1,-1,-1,-1)$ and the Lorentz scalar  product which is defined by $(a.b)=a^\mu b_\mu$. The layout of this paper is as follows : the presentation of the necessary formalism of this work in section [2 and 3], the result and discussion in section 4 and at last a brief conclusion in section 5.
\section{Theory of the inelastic collision in the absence of de laser field}
In this section, we calculate the exact analytical expression of the
semirelativistic unpolarized DCS for the relativistic excitation of
atomic hydrogen by electron impact. The transition matrix element
for the direct channel (exchange effects are neglected) is given by
\begin{eqnarray}
S_{fi} &=&-i\int dt\langle\psi _{p_{f}}(x_{1})\phi _{f}(x_{2})\mid
V_{d}\mid \psi _{p_{i}}(x_{1})\phi _{i}(x_{2})\rangle
 \nonumber \\
 &=&-i\int_{-\infty }^{+\infty }dt\int d\mathbf{r}_{1}\overline{\psi }
_{p_{f}}(t,\mathbf{r}_{1})\gamma^{0}\psi _{p_{i}}(t,\mathbf{r}_{1})
\langle\phi _{f}(x_{2})\mid V_{d}\mid \phi _{i}(x_{2})\rangle
\label{1}
\end{eqnarray}
where
\begin{equation}
V_{d}=\frac{1}{r_{12}}-\frac{Z}{r_{1}}  \label{2}
\end{equation}
is the direct interaction potential, $\mathbf{r}_{1}$ are the
coordinates of the incident and scattered electron, $\mathbf{r}_{2}$
 the atomic electron coordinates, $r_{12}=$ $\mid
\mathbf{r}_{1}-\mathbf{r}_{2}\mid $ and $r_{1}=\mid
\mathbf{r}_{1}\mid $. The function $\psi _{p_{i}}(x_{1})=\psi
_{p}(t,\mathbf{r}_{1})=u(p,s)\exp (-ip.x)/\sqrt{2EV}$ is the
electron wave function, described by a free Dirac spinor normalized
to the volume $V$, and $\phi _{i,f}(x_{2})=\phi
_{i,f}(t,\mathbf{r}_{2})$ are the semirelativistic wave functions of
the hydrogen atom where the index $i$ and $f$ stand for the initial
and final states respectively. The semirelativistic wave function of
the atomic hydrogen is the Darwin wave function for bound states
\cite{17}, which is given by :
\begin{equation}
\phi _{i}(t,\mathbf{r}_{2})=\exp (-i\mathcal{E} _{b}(1s_{1/2})t)\varphi_{1s} ^{(\pm )}(%
\mathbf{r}_{2})  \label{3}
\end{equation}
where $\mathcal{E} _{b}(1s_{1/2})$ is the binding energy of the ground
state of atomic hydrogen and $\varphi_{1s} ^{(\pm
)}(\mathbf{r}_{2})$ is  given by :
\begin{equation}
\varphi_{1s} ^{(\pm
)}(\mathbf{r}_{2})=(\mathsf{1}_{4}-\frac{i}{2c}\mathbf{\alpha
.\nabla }_{(2)})u^{(\pm )}\varphi _{0}(\mathbf{r}_{2})  \label{4}
\end{equation}
it represents a quasi relativistic bound state wave function,
accurate to first order in $Z/c$ in the relativistic corrections
(and normalized to the same order), with $\varphi _{0}$ being the
non-relativistic bound state hydrogenic function. The spinors
$u^{(\pm )}$ are such that $u^{(+)}=(1,0,0,0)^{T}$ and
$u^{(-)}=(0,1,0,0)^{T}$ and represent the basic four-component
spinors for a particle at rest with spin-up and spin-down,
respectively. The matrix differential operator $\alpha.\nabla$ is
given by :
\begin{eqnarray}
 \alpha.\nabla=\left(
 \begin{array}{cccc}
     0 & 0 & \partial_z & \partial_x-i\partial_y \\
  0 & 0 & \partial_x+i\partial_y & -\partial_z \\
  \partial_z & \partial_x-i\partial_y & 0 & 0 \\
 \partial_x+i\partial_y &- \partial_z & 0 & 0
  \end{array}
\right)
\end{eqnarray}
For the spin up, we have :
\begin{eqnarray}
\varphi_{1s} ^{(+)}(\mathbf{r}_{2})&=&N_{D_1} \left(
\begin{array}{c}
 1 \\
  0 \\
  \frac{i}{2cr_2}z \\
  \frac{i}{2cr_2}(x+iy)
\end{array}\right)
\frac{1} {\sqrt{\pi }} e^{-r_{2}}
\end{eqnarray}
and for the spin down, we have :
\begin{eqnarray}
\varphi_{1s} ^{(-)}(\mathbf{r}_{2})&=& N_{D_1}\left(
\begin{array}{c}
 0 \\
  1 \\
  \frac{i}{2cr_2}(x-iy) \\
  -\frac{i}{2cr_2}z
\end{array}\right)
\frac{1} {\sqrt{\pi }} e^{-r_{2}}
\end{eqnarray}
where
\begin{equation}
N_{D_1}=2c/\sqrt{4c^{2}+1}  \label{7}
\end{equation}
is a normalization constant lower but very close to 1. Let us
mention that the function $\phi _{f}(t,\mathbf{r}_{2})$ in Eq. (1)
is the Darwin wave function for bound states \cite{18}, which is also
accurate to the order $Z/c$ in the relativistic corrections. This is
expressed as $\phi
_{f}(t,\mathbf{r}_{2})=\exp (-i\mathcal{E}_{b}(2s_{1/2})t)\varphi_{2s} ^{(\pm )}(%
\mathbf{r}_{2})$ with $\mathcal{E}_{b}(2s_{1/2})$ as the binding energy
of the $2s_{1/2}$ state of atomic hydrogen.
\begin{eqnarray}
\varphi_{2s} ^{(+)}(\mathbf{r}_{2})&=&N_{D_2} \left(
\begin{array}{c}
 2-r_2 \\
  0 \\
  \frac{i(4-r_2)}{4r_2c}z \\
  \frac{(4-r_2)}{4rc}(-y+ix)
\end{array}\right)
\frac{1} {4\sqrt{2\pi }} e^{-r_{2}}
\end{eqnarray}
 for the spin up and
\begin{eqnarray}
\varphi_{2s} ^{(-)}(\mathbf{r}_{2})&=& N_{D_2}\left(
\begin{array}{c}
 0 \\
  2-r_2 \\
  \frac{4-r_2}{4cr_2}(y+ix)\\
  i\frac{(r_2-4)}{4cr_2}z
\end{array}\right)
\frac{1} {4\sqrt{2\pi }} e^{-r_{2}}
\end{eqnarray}
for the spin down. The transition matrix element in Eq. (\ref{1})
becomes :
\begin{eqnarray}
S_{fi}&=&-i\int_{-\infty }^{+\infty }dt\int d\mathbf{r}_{1}
d\mathbf{r}_{2}\overline{\psi }
_{p_{f}}(t,\mathbf{r}_{1})\gamma^{0}\psi
_{p_{i}}(t,\mathbf{r}_{1}) \phi^{\dag} _{f}(t,r_{2})\phi
_{i}(t,r_{2}) V_{d} \label{10}
\end{eqnarray}
and it is straightforward to get, for the transition amplitude,
\begin{eqnarray}
S_{fi} &=&-i\frac{\overline{u}(p_f,s_f)\gamma^0
u(p_i,s_i)}{2V\sqrt{E_fE_i}}2\pi H_{inel}(\Delta)
\delta\big(E_f+\mathcal{E}(2s_{1/2})-E_i-\mathcal{E}(1s_{1/2})\big)\nonumber\\
\label{11}
\end{eqnarray}
where  $\Delta=|p_i-p_f|$ and $\gamma^0$ is the Dirac matrix.
Using the  standard technique of the QED, we find for the
unpolarized DCS
\begin{eqnarray}
\frac{d\overline{\sigma}}{d\Omega_f}&=&\frac{|\mathbf{p}_f|}{|\mathbf{p}_i|}\frac{1}{(4\pi
c^2)^2}\left(\frac{1}{2}\sum_{s_is_f}|\overline{u}(p_f,s_f)\gamma^0
u(p_i,s_i)|^2\right)\left|H_{inel}(\Delta)\right|^2 \label{12}
\end{eqnarray}
\section{Calculation of the integral part}
The function $H_{inel}(\Delta)$ is found if one performs the various
integrals :
\begin{eqnarray}
H_{inel}(\Delta)=\int_0^{+\infty}d\mathbf{r}_1 e^{i\mathbf{\Delta}
\mathbf{r}_1}I(\mathbf{r}_1)\label{13}
\end{eqnarray}
%\subsection{Integral over $\mathbf{r}_2$}
The quantity $I(\mathbf{r}_1)$ is easily evaluated in the following
way. We first write the explicit form of $I(\mathbf{r}_1)$ :
\begin{eqnarray}
I(\mathbf{r}_1)=\int_0^{+\infty}d\mathbf{r}_2\phi^{\dag}
_{2s}(\mathbf{r}_{2})\left[\frac{1}{r_{12}}-\frac{Z}{r_1}\right]\phi _{1s}(\mathbf{r}_{2})
\label{14}
\end{eqnarray}
Next, we develop the quantity $r^{-1}_{12}$ in spherical harmonics
as
\begin{eqnarray}
\frac{1}{\mathbf{r}_{12}}=4\pi\sum_{lm}\frac{Y_{lm}(\widehat{r}_1)Y_{lm}^*(\widehat{r}_2)}{2l+1}\frac{(\mathbf{r}_<)^l}{(\mathbf{r}_>)^{l+1}}\label{15}
\end{eqnarray}
where $r_>$ is the greater of $r_1$ and $r_2$, and $r_<$ the lesser
of them. The angular coordinates of the vectors $\mathbf{r}_1$ and
$\mathbf{r}_2$ are such that : $\widehat{r}_1=(\theta_1,\varphi_1)$
and $\widehat{r}_2=(\theta_2,\varphi_2)$. We use the well known
integral \cite{19}
\begin{eqnarray}
\int_x^{+\infty} du \;u^{m} e^{-\alpha u}= \frac{m!}{\alpha^{m+1}} e^{-\alpha x}\sum_{\mu=0}^m \frac{\alpha^\mu x^\mu}{\mu ! }\qquad \qquad Re(\alpha)>0 \label{16}
\end{eqnarray}
then, after some analytic calculations, we get for $I(\mathbf{r}_1)$
:
\begin{eqnarray}
I(\mathbf{r}_1)=\frac{6}{27}(\frac{1}{c^2}-4)+\frac{4}{27c^2}\frac{1}{\mathbf{r}_1}-\frac{4}{9}(1+\frac{1}{8c^2})\mathbf{r}_1\label{17}
\end{eqnarray}
%\subsection{Integral over $\mathbf{r}_1$}
The integration over $\mathbf{r}_1$ gives rise to the following formula :
\begin{eqnarray}
H_{inel}(\Delta)=\int_0^{+\infty}d\mathbf{r}_1 e^{i\mathbf{\Delta} \mathbf{r}_1} I(\mathbf{r}_1)=-\frac{4\pi}{\sqrt{2}}(I_1+I_2+I_3)
\label{18}
\end{eqnarray}
the angular integrals are performed by expanding the plane wave
$e^{i\mathbf{\Delta} \mathbf{r}_1}$ in spherical harmonics as :
\begin{eqnarray}
e^{i\mathbf{\Delta} \mathbf{r}_1} = \sum_{lm}4\pi i^l
j_l(\mathbf{\Delta}
\mathbf{r}_1)Y_{lm}(\widehat{\mathbf{\Delta}})Y_{lm}^*(\widehat{\mathbf{r}}_1)
\label{19}
\end{eqnarray}
with $\mathbf{\Delta}=\mathbf{p}_i-\mathbf{p}_f$ is the relativistic
momentum transfer and $\widehat{\mathbf{\Delta}}$ is the angular
coordinates of the vector $\mathbf{\Delta}$. Then, after some
analytic computations, we get for $I_1$, $I_2$ and $I_3$ the
following result :
\begin{eqnarray}
I_1&=&\frac{4}{27c^2}\int_0^{+\infty}dr_1\;r_1e^{-\frac{3}{2}r_1}j_0(\Delta r_1)=\frac{4}{27c^2}\frac{1}{((3/2)^2+\mathbf{\Delta}^2)}\nonumber\\
I_2&=&\frac{6}{27}(\frac{1}{c^2}-4)\int_0^{+\infty}dr_1\;r_1^2e^{-\frac{3}{2}r_1}j_0(\Delta r_1)=\frac{2}{27}(\frac{1}{c^2}-4)\frac{3}{((3/2)^2+\mathbf{\Delta}^2)^2}\\
\label{13}
I_3&=&-\frac{4}{9}(1+\frac{1}{8c^2})\int_0^{+\infty}dr_1\;r_1^3e^{-\frac{3}{2}r_1}j_0(\Delta
r_1)=\frac{8}{9}(1+\frac{1}{8c^2})\frac{\mathbf{\Delta}^2-27/4}{((3/2)^2+\mathbf{\Delta}^2)^3}\nonumber
\end{eqnarray}
It is clear that the situation is different than in elastic
collision \cite{14}, since we have no singularity in the case
($\mathbf{\Delta} \to 0$)

%\section{Calculation of the spinorial part}
The calculation the spinorial part is reduced to the computation of traces of
$\gamma$ matrices. This is routinely done using Reduce \cite{20}. We
consider the unpolarized DCS. Therefore, the various polarization
states have the same probability and the actual calculated spinorial part is given by summing over the final
polarization $s_f$ and averaging aver the initial polariztion $s_i$. Therfore, the spinorial
part is given by :
\begin{eqnarray}
\frac{1}{2}\sum_{s_is_f}|\overline{u}(p_f,s_f)\gamma^0
u(p_i,s_i)|^2&=&\textbf{Tr}\left\{\gamma^0(\ps_ic+c^2)\gamma^0(\ps_fc+c^2)\right\}\nonumber\\
&=&2c^2[\frac{2E_fE_i}{c^2}-(p_i.p_f)+c^2]\label{14}
\end{eqnarray}
We must, of course, recover the result in the nonrelativistic limit
($\gamma \longrightarrow 1$), situation of which the differential
cross section can simply given by :
\begin{eqnarray}
\frac{d\overline{\sigma}}{d\Omega_f}=\frac{|\mathbf{K}_f|}{|\mathbf{K}_i|}\frac{128}{\left(|\mathbf{\Delta_{nr}}|^2+\frac{9}{4}\right)^6}\label{14}
\end{eqnarray}
with $|\mathbf{\Delta_{nr}}|=|\mathbf{K}_i-\mathbf{K}_f|$ is the
nonrelativistic momentum transfer and the momentum vectors
($\mathbf{K}_i$, $\mathbf{K}_f$) are related by the following
formula :
\begin{eqnarray}
\mathbf{K}_f=(|\mathbf{K}_i|^2-3/4)^{1/2}
\end{eqnarray}
\section{Theory of the inelastic collision in the presence of de laser field}
The second-order Dirac equation for the elctron in the
presence of an external electromagnetic field is given by :
\begin{eqnarray}
\left[\left(p-\frac{1}{c}A\right)^2-c^2-\frac{i}{2c}F_{\mu\nu}\sigma^{\mu\nu}\right]\psi(x)=0\label{15}
\end{eqnarray}
where  $\sigma^{\mu\nu}=\frac{1}{2}[\gamma^\mu,\gamma^\nu]$ is the tensors related to Dirac matrices $\gamma^\mu$ and $F^{\mu\nu}=\partial_\mu
A_\nu-\partial_\nu A_\mu $ is the electromagnetic field tensor. $A^\mu$ is the
four-vector potential. The plane wave solution of the second-order equation is known as the
Volkov state \cite{13}.
\begin{eqnarray}
\psi(x)=\left(1+\frac{\ks\As}{2c(kp)}\right)\frac{u(p,s)}{\sqrt{2VQ_0}}\exp\left[-i(qx)-i\int_0^{kx}\frac{(Ap)}{c(kp)}d\phi\right]\label{16}
\end{eqnarray}
We turn now to the calculation of the laser-assisted transition amplitude. The instantaneous interaction potential is given by
\begin{eqnarray}
V_d=\frac{1}{\mathbf{r}_{12}}-\frac{Z}{\mathbf{r}_{1}}\label{17}
\end{eqnarray}
where $\mathbf{r}_{12}=|\mathbf{r}_{1}-\mathbf{r}_{2}|$; $\mathbf{r}_{1}$ are the electron coordinates and $\mathbf{r}_{2}$ are the atomic coordinates. The transition matrix
element corresponding to the process of laser assisted electron-atomic
hydrogen from the initial state $i$ to the final state $f$ is given by
\begin{eqnarray}
S_{fi}=-i\int dt\; \langle\overline{\psi}_{q_f}(\mathbf{r}_1)\phi_f(\mathbf{r}_2)|V_d| \psi_{q_i}(\mathbf{r}_1)\phi_i(\mathbf{r}_2)\rangle\label{18}
\end{eqnarray}
Proceeding along the lines of standard calculations in QED, one has for the DCS
\begin{eqnarray}
\frac{d\overline{\sigma}}{d\Omega_f}=\sum_{s=-\infty}^{+\infty}\frac{d\overline{\sigma}^{(s)}}{d\Omega_f}\label{19}
\end{eqnarray}
with
\begin{eqnarray}
\frac{d\overline{\sigma}^{(s)}}{d\Omega_f}=\left.\frac{|\mathbf{q}_f|}{|\mathbf{q}_i|}\frac{1}{(4\pi
c^2)^2}\left(\frac{1}{2}\sum_{s_is_f}|M_{fi}^{(s)}|^2\right)\left|H_{inel}(\Delta_s)\right|^2\right|_{Q_f=Q_i+s\omega+E_{1s1/2}-E_{2s1/2}}. \label{20}
\end{eqnarray}
The novelty in the various stages of the calculations
is contained in the spinorial part $\frac{1}{2}\sum_{s_i}\sum_{s_f}|M_{fi}^n|^2$ that contains all the information about the
spin and the laser-interaction effects. This quantity can be obtained using REDUCE \cite{20} and are explicitly given in our previous work \cite{20'}. After some analytical calculations, the integral part $H_{inel}(\Delta_s)$ reduces to
\begin{eqnarray}
H_{inel}(\Delta_s)=-\frac{4\pi}{\sqrt{2}}[I_1(s)+I_2(s)+I_3(s)]\label{22}
\end{eqnarray}
with
\begin{eqnarray}
I_1(s)&=&\frac{4}{27c^2}\frac{1}{((3/2)^2+\mathbf{\Delta_s}^2)}\nonumber\\
I_2(s)&=&\frac{2}{27}(\frac{1}{c^2}-4)\frac{3}{((3/2)^2+\mathbf{\Delta_s}^2)^2}\\
\label{23}
I_3(s)&=&\frac{8}{9}(1+\frac{1}{8c^2})\frac{\mathbf{\Delta_s}^2-27/4}{((3/2)^2+\mathbf{\Delta_s}^2)^3}\nonumber
\end{eqnarray}
with $\mathbf{\Delta}_s=\mathbf{q}_f-\mathbf{q}_i-s\mathbf{k}$ is the momentum transfer in the presence of the laser field.
\section{Results and discussions}
\subsection{In the absence of the laser field}
\begin{figure}[!h]
 \centering
\includegraphics[angle=0,width=3 in,height=3.5 in]{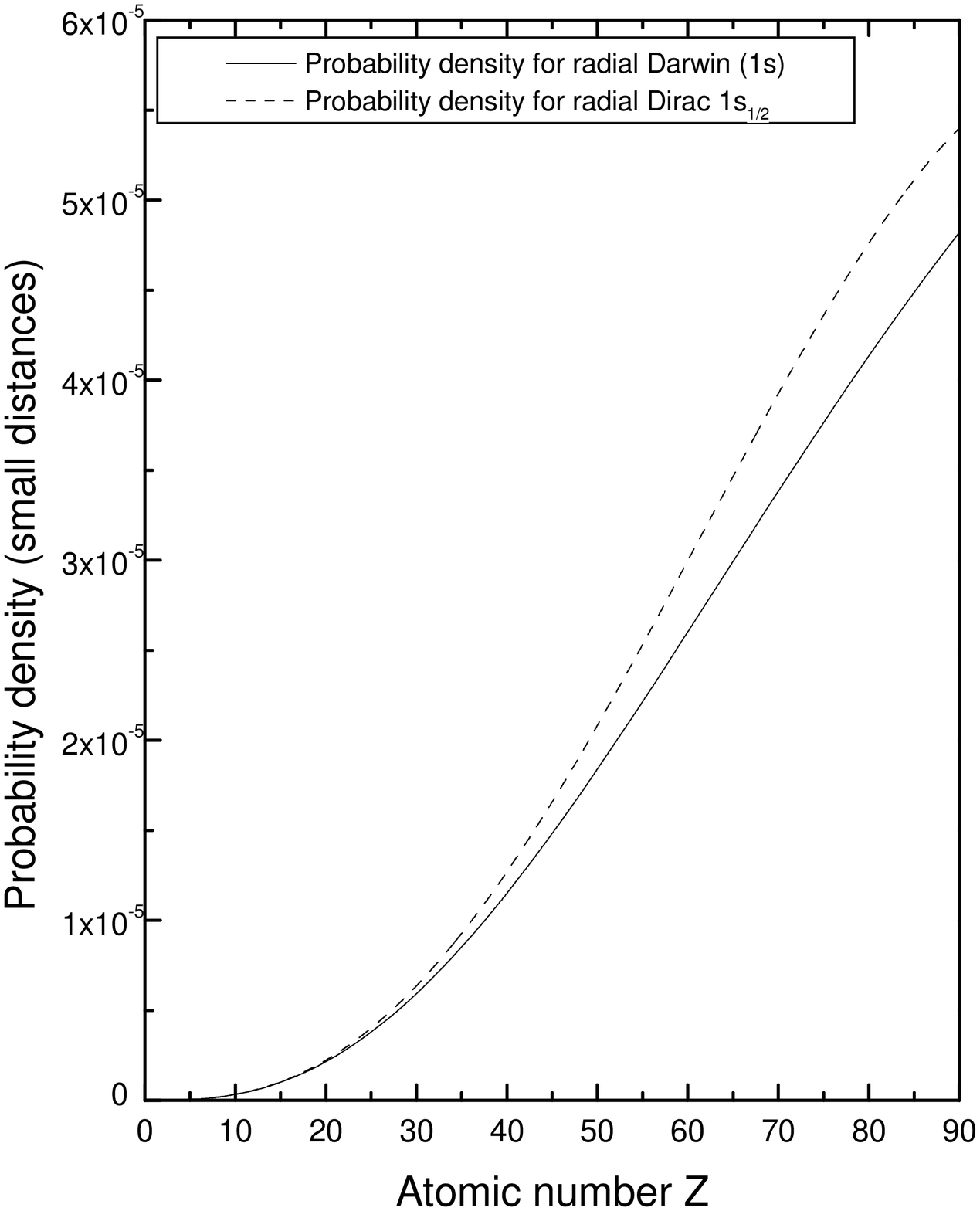}
\caption{\label{fig1}Behavior of the probability density for  radial
Darwin wave function compared with that of the Dirac wave function
for small distances and for increasing values of the atomic charge
number.}
\end{figure}
In presenting our results it is convenient to consider separately
those corresponding to non-relativistic regime (the relativistic
parameter $\gamma \simeq 1$) and those related to relativistic one
(the relativistic parameter $\gamma \simeq 2$). Before beginning the
discussion of the obtained results, it is worthwhile to recall the
meaning of some abbreviation that will appear throughout this
section. The NRDCS stands for the nonrelativistic differential cross
section, where nonrelativistic plane wave are used to describe the
incident and scattered electrons. The SRDCS stands for the
semirelativistic differential cross section.\\
\indent We begin our numerical work with the study of the dependence
of the probability density for radial Darwin and Dirac wave
functions on the atomic charge number $Z$.
\begin{figure}[!h]
 \begin{minipage}[b]{.46\linewidth}
  \centering\epsfig{figure=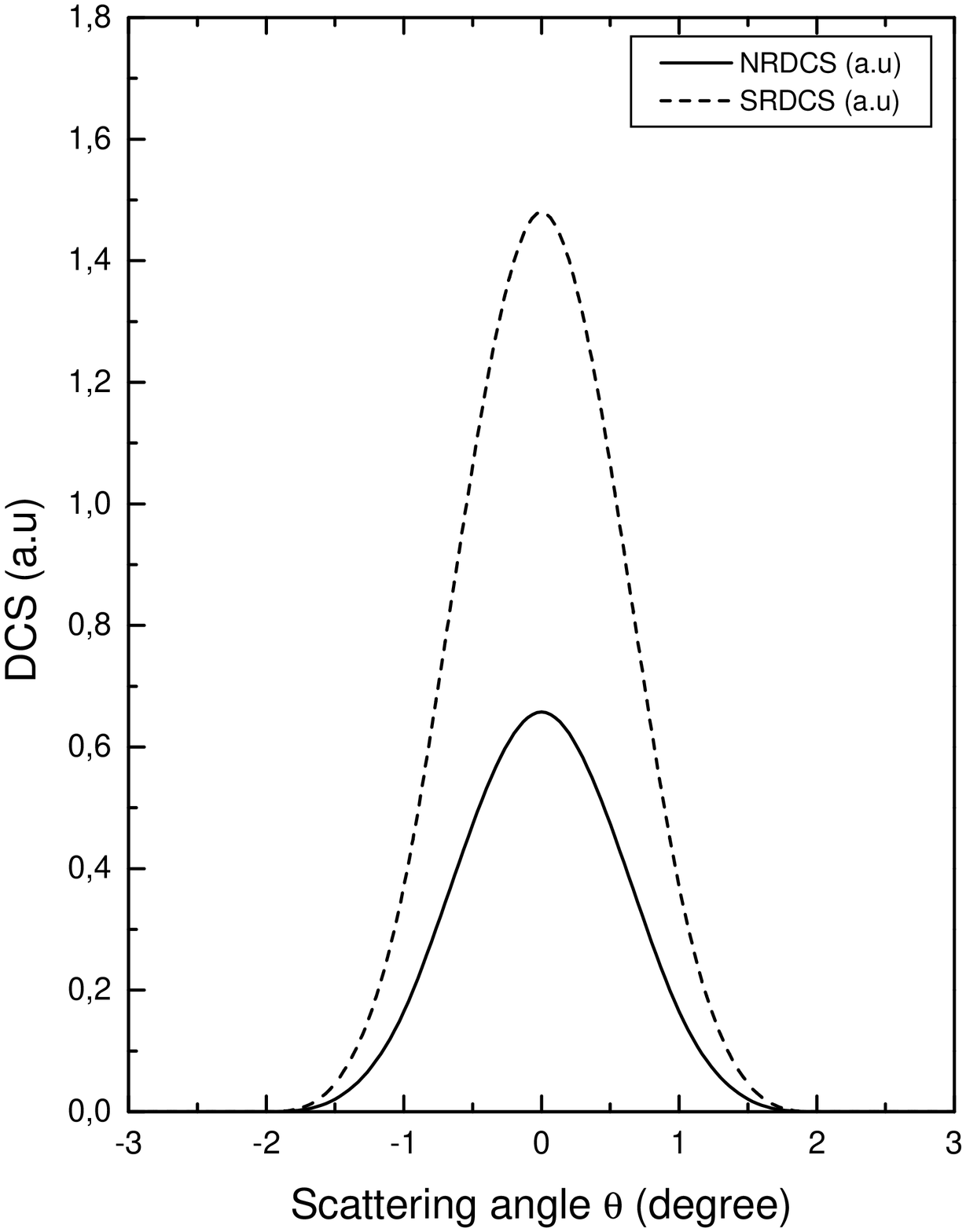,width=\linewidth,height=3.3in}
  \caption{\small{ \label{fig2} The long-dashed line represents the
semi-relativistic DCS, the  solid line represents the corresponding
non-relativistic DCS for a relativistic parameter ($\gamma=1.5$) as
functions of the scattering angle $\theta $. \vspace{0.4 cm} }}
 \end{minipage} \hfill
 \begin{minipage}[b]{.46\linewidth}
  \centering\epsfig{figure=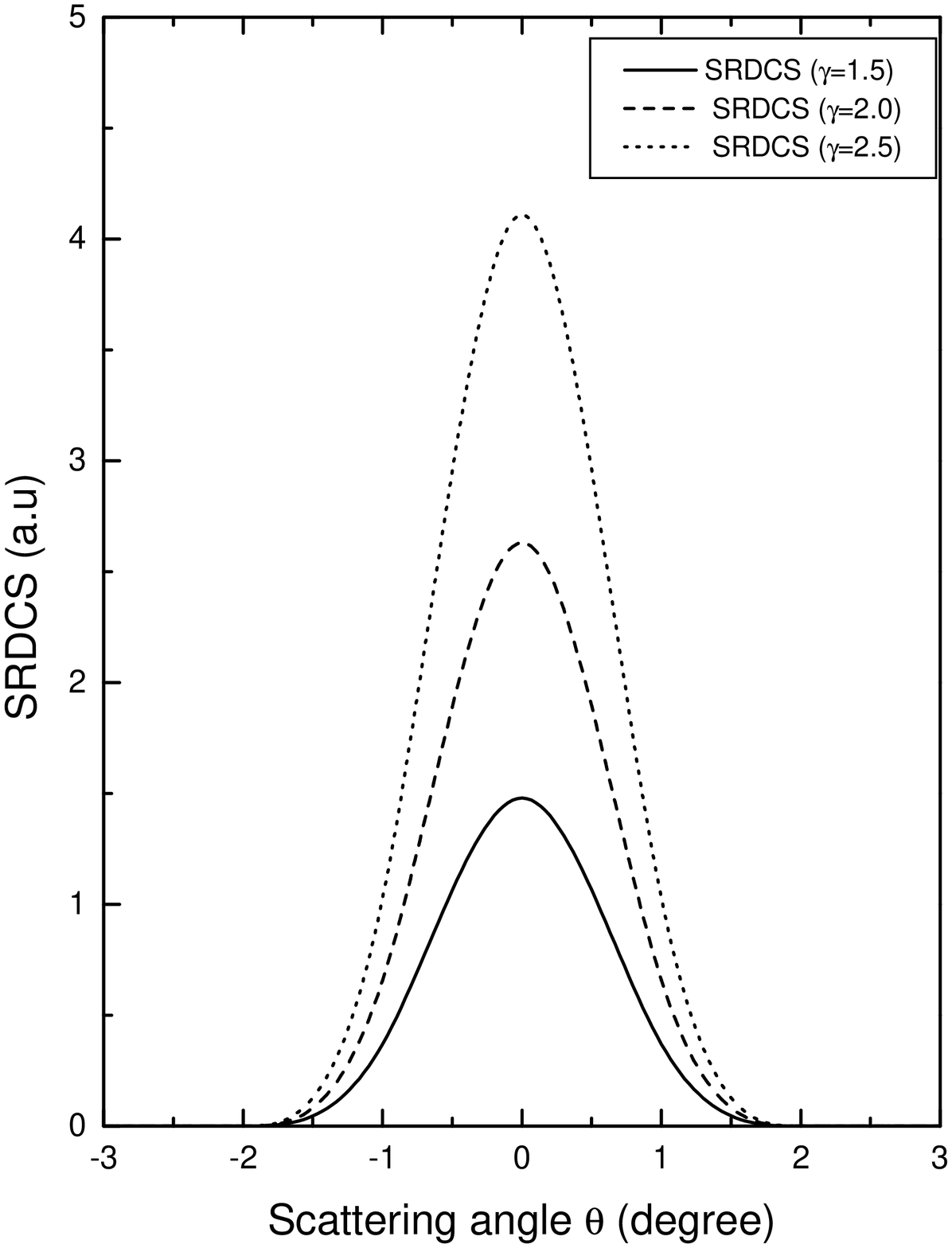,width=\linewidth,height=3.3in}
  \caption{\small{ \label{fig3} The solid line represents the
semi-relativistic DCS, the long-dashed line represents the
corresponding non-relativistic DCS  for various values of the
relativistic parameter ($\gamma=1.5$, $\gamma=2$ and $\gamma=2.5$)
as functions of the scattering angle $\theta$.}}
 \end{minipage}
\end{figure}

\begin{figure}[!h]
 \begin{minipage}[b]{.46\linewidth}
  \centering\epsfig{figure=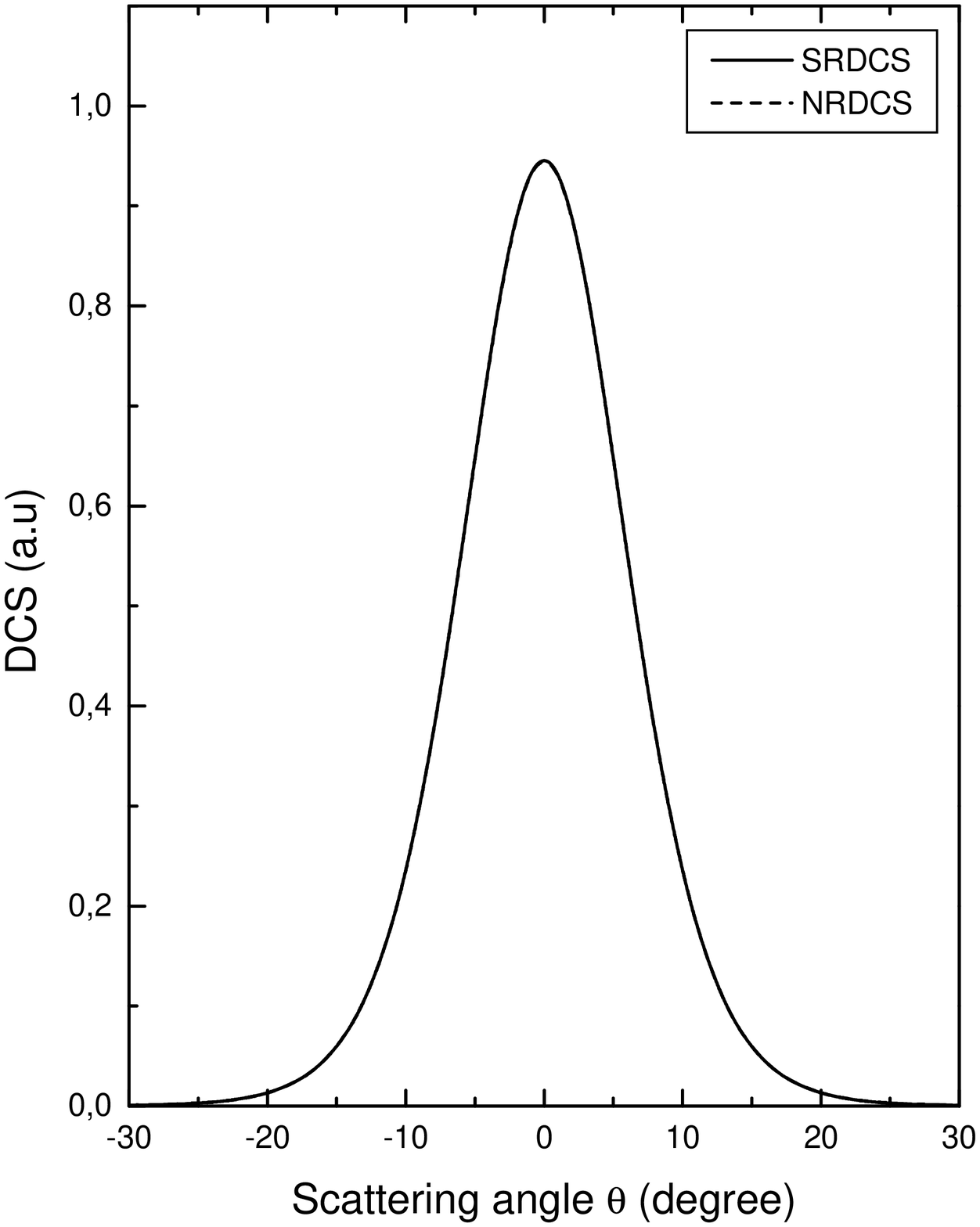,width=\linewidth,height=3.3in}
  \caption{\small{\label{fig4} The solid line represents the
semi-relativistic DCS, the long-dashed line represents the
corresponding non-relativistic DCS for a relativistic parameter
$\gamma=1.00053$ as functions of the scattering angle
$\theta$.\vspace{0.4cm}}}
 \end{minipage} \hfill
 \begin{minipage}[b]{.46\linewidth}
  \centering\epsfig{figure=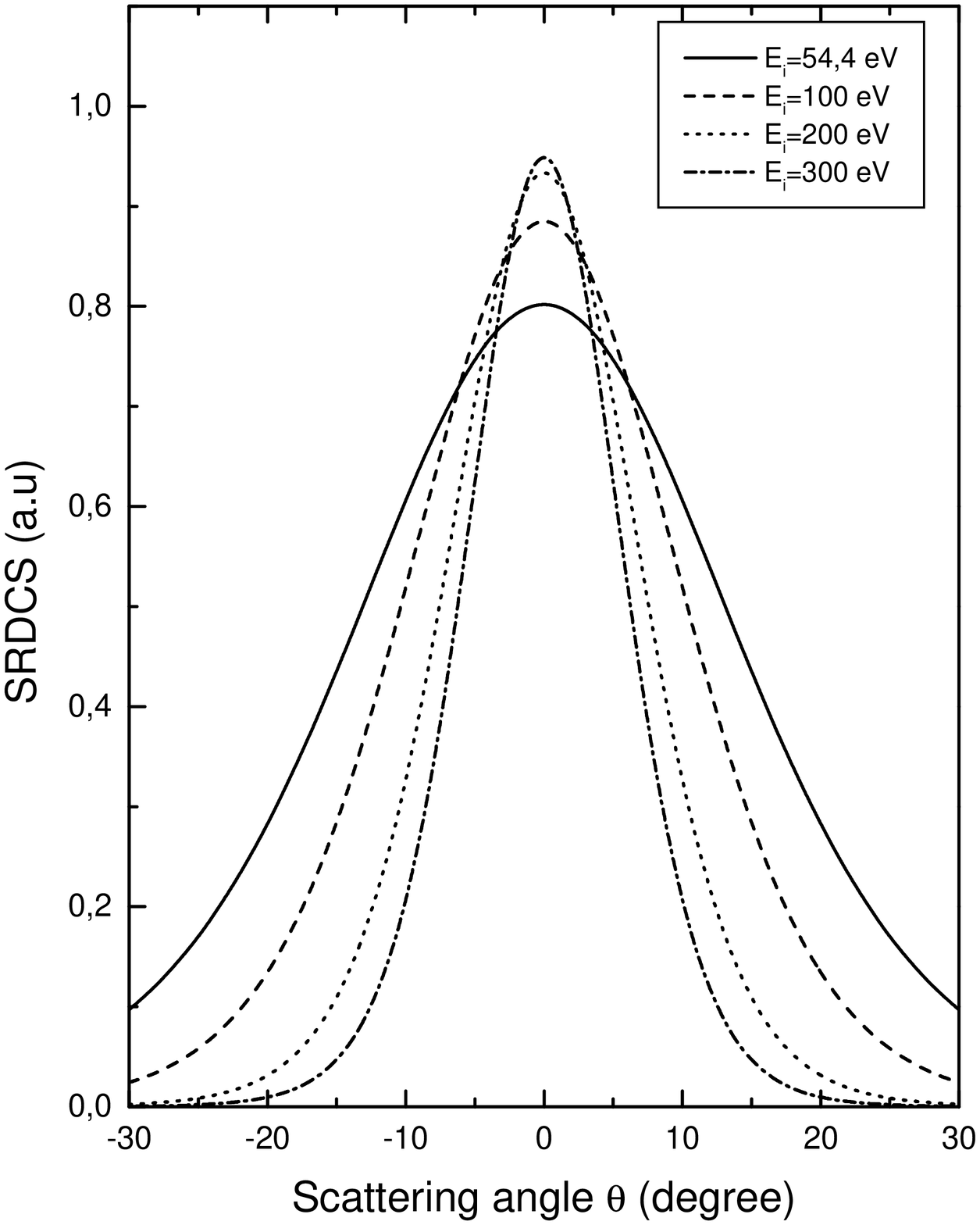,width=\linewidth,height=3.3in}
  \caption{\small{ \label{fig5} The variation of the SRDCSs with respect to $\theta$,
   for various kinetic energies.\vspace{1.8cm}}}
 \end{minipage}
 \end{figure}
So long as the condition $Z\alpha \ll 1$ is verified, the use of
Darwin wave function do not have any influence at all on the results
at least in the first order of perturbation theory. So, the
semirelativistic treatment when $Z$ increases may generate large
errors but not in the case of this work. In this paper, we can not
have numerical instabilities since there are none. For the sake of
illustration, we give in figure 1 the behavior of the probability
density for radial Darwin wave functions as well as that of the
exact relativistic Dirac wave functions for different values of $Z$.
As you may see, even if it is not noticeable in figure 1, there are
growing discrepancies for $Z=10$ and these become more pronounced
when $Z=20$. The QED formulation shows that there are relativistic
and spin effects at the relativistic domain and the non relativistic
formulation is no
longer valid.\\
\begin{figure}[!h]
 \centering
\includegraphics[angle=0,width=3 in,height=3.3 in]{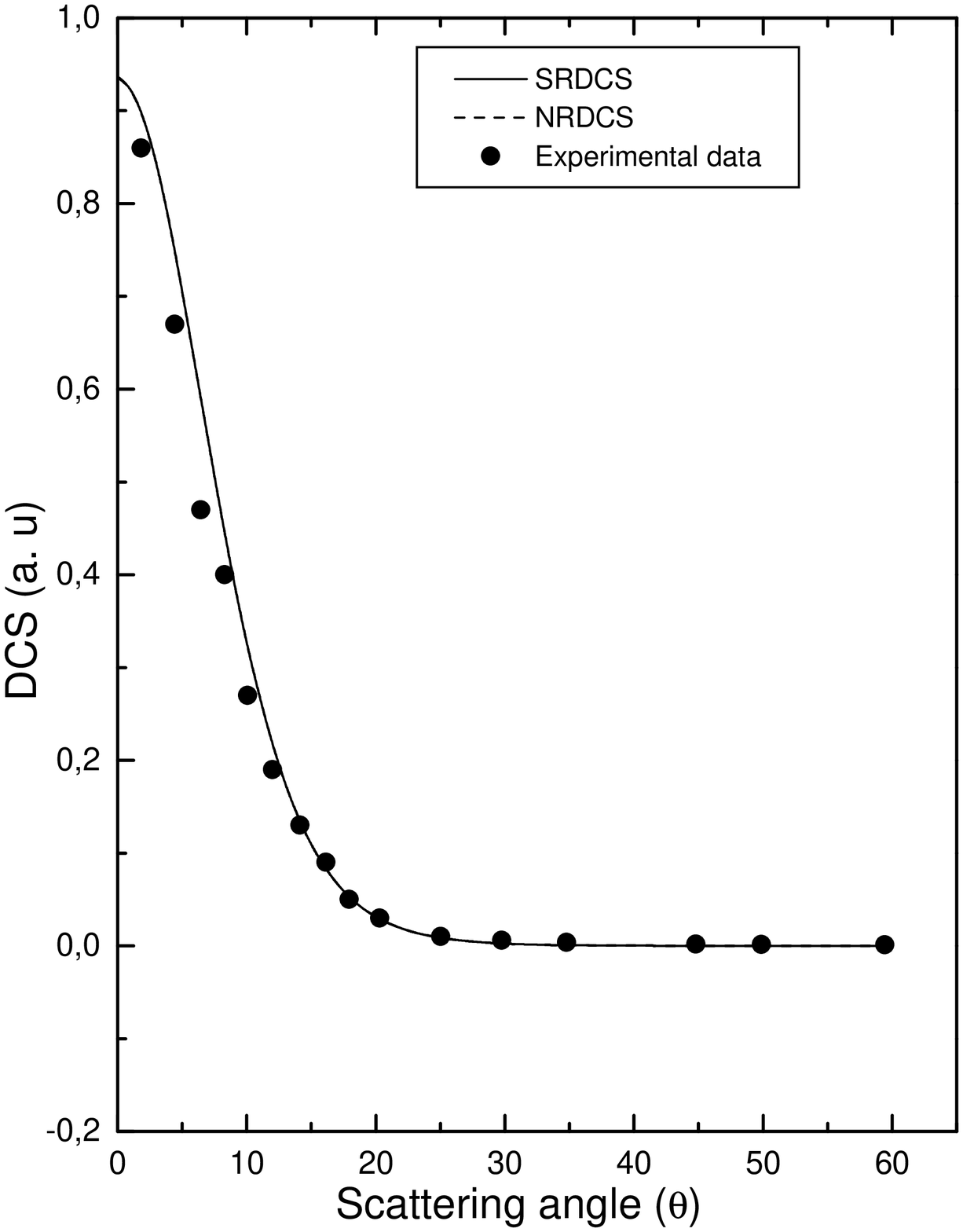}
\caption{\label{fig6} The variation of the differential $1s-2s$
cross section of $e^--H$ scattering at $200\; eV$. The dots are the
observed values of J. F. Wiliams (1981) ; the solid line represents
the semi-relativistic approximation and the long-dashed line
corresponds to the non-relativistic DCS.}
\end{figure}
\indent In the relativistic regime, the semirelativistic
differential cross section results obtained for the
$1s\longrightarrow 2s$ transition in atomic hydrogen by electron
impact, are displayed in figures 2 and 3. In this regime, there are
no theoretical models and experimental data for comparison as in
nonrelativistic regime. In such a situation, it appears from figures
\ref{fig2} and \ref{fig3} that in the limit of high electron kinetic
energy, the effects of the additional spin terms and the relativity
begin to be noticeable and that the non-relativistic formalism is no
longer applicable. Also a peak in the vicinity of
$\theta_f=0^{\circ}$ is clearly observed.\\
\indent The investigation in the nonrelativistic regime was carried
out with $\gamma$ as a relativistic parameter and $\theta$ as a
scattering angle. In atomic units, the kinetic energy is related to
$\gamma$ by the following relation : $E_k=c^2(\gamma-1)$. Figure 4
shows the dependence of DCS, obtained in two models (SRDCS, NRDCS),
on scattering angle $\theta$.  In this regime, it appears clearly
that there is no difference between these models. Figure 5 shows the
variation of the SRDCS with $\theta$ for various energies. It also
shows approximatively in the interval [-5, 5] that the SRDCS
increases with $\gamma$, but decreases elsewhere. Figure 6 presents
the observed and calculated angular dependence of $1s-2s$
differential cross section of $e^--H$ scattering at incident energie
$200\;eV$. Results obtained in the approaches (semirelativistic and
nonrelativistic approximations) are indistinguishable and in good
agreement with the experimental data provided by J. F. Williams
\cite{21}.
\subsection{In the presence of the laser field}
\begin{figure}[!t]
 \centering
\includegraphics[angle=0,width=3 in,height=3.3 in]{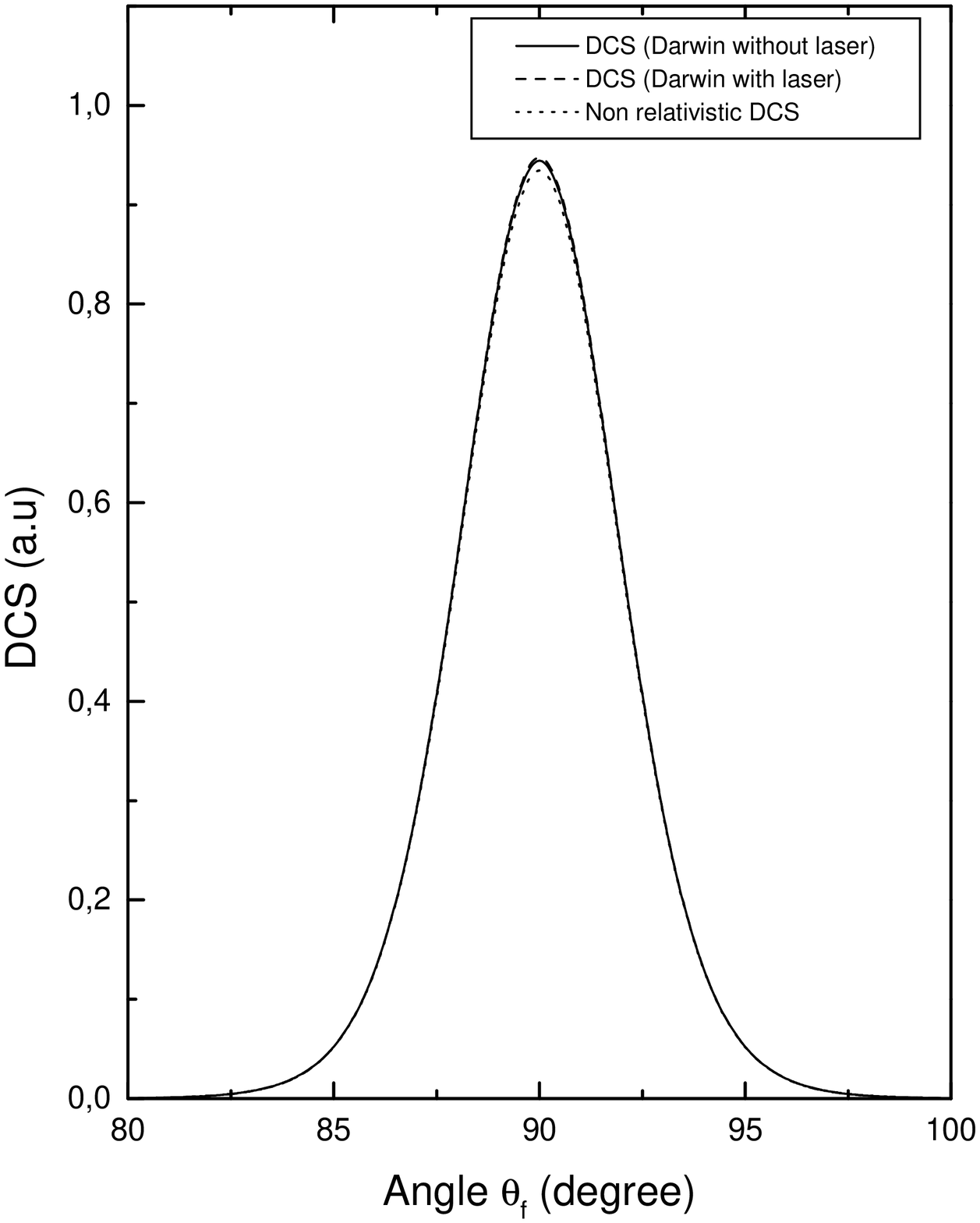}
\caption{\label{fig7} The various DCSs as a function of $\theta_f$ for an electrical field strength of $\mathcal{E}=0 \; a.u$, a relativistic parameter $\gamma=1.0053$}
\end{figure}

\begin{figure}[!h]
 \begin{minipage}[b]{.46\linewidth}
  \centering\epsfig{figure=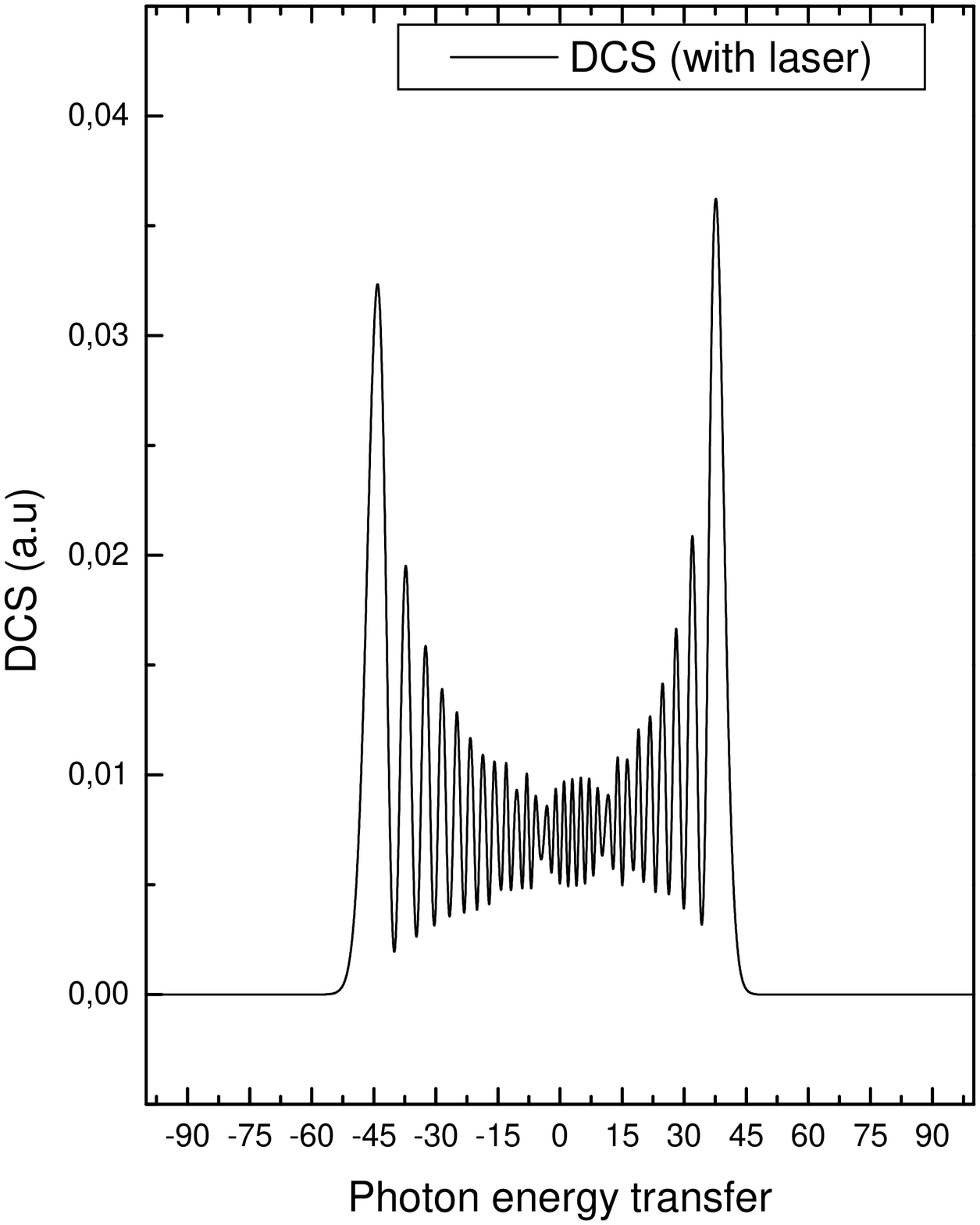,width=\linewidth,height=3.3in}
  \caption{\small{\label{fig8} Envelope of the SRDCS in the non relativistic regime ($\gamma=1.0053$ and $\mathcal{E}=0.05\;a.u$) for a number of net photons exchanged $\pm 100 $.}}
 \end{minipage} \hfill
 \begin{minipage}[b]{.46\linewidth}
  \centering\epsfig{figure=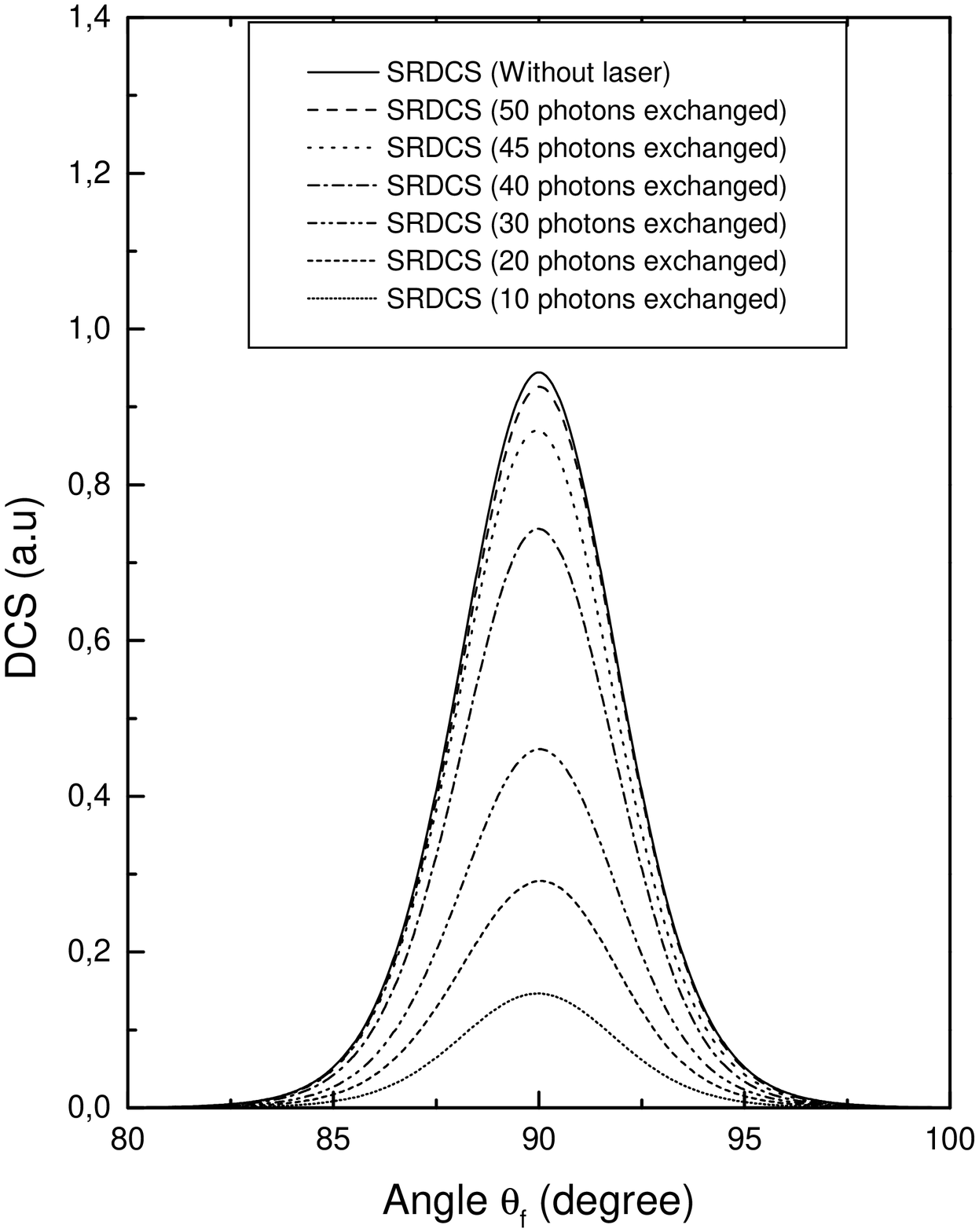,width=\linewidth,height=3.3in}
  \caption{\small{ \label{fig9} The variation of the SRDCSs with and without laser
   for various numbers of photons exchanged ($\pm 10$, $\pm 20$, $\pm 20$, $\pm 40$, $\pm 45$) and $\pm 50$).}}
 \end{minipage}
 \end{figure}

For the description of the scattering geometry, we work in a coordinate system in which $k||\widehat{e}_z$. This means that the direction of the laser propagation is along $Oz$ axis. We begin by defining our scattering geometry. The undressed angular coordinates of the incoming electron are $\theta_i=90^{\circ}$, $\phi_i=45^{\circ}$ and for the scattered electron, we have $0^{\circ}\leq \theta_f\leq 180^{\circ}$ and $\phi_f=45^{\circ}$. When we take the zero electric field strength ($\mathcal{E}=0 \; a.u$), one finds overlapping curves for the tree approaches [SRDCS (without laser), SRDCS (with laser) and the Non relativistic DCS]. It represents our first consistency check of our calculations and it is shown in figure \ref{fig7}.
Figure \ref{fig8} shows the envelope of the laser assisted SRDCS as a function of the net number of photons exchanged. The cutoffs are $s\simeq -50 $ photons for the negative part of the envelope and $s\simeq +50 $ photons for the positive part. In figure \ref{fig9}, for the geometry $\theta_i=90^{\circ}$, $\phi_i=45^{\circ}$ and $\phi_f=45^{\circ}$, we have made simulations concerning the laser-assisted SRDCS for a set of net number photons exchanged. These sets ($\pm 10$, $\pm 20$, $\pm 30$, $\pm 40$, $\pm 45$ and $\pm 50$) show that at $\pm 50$ the SRDCS (with laser) is almost close to the SRDCS (without laser).
 But for instance at $\pm 10$, we have several orders of magnitude as a result of the difference between the two approaches. We return to the first case $\pm 50$, the convergence reached here is called the sum-rule that was shown by Bunkin and Fedorov as well as by Kroll and Watson \cite{22}. The figures (\ref{fig8} and \ref{fig9}) correctly introduce correlation in the net number of photons exchanged that reaches the well-known sum-rule. As you can see from the figure \ref{fig8}, the SRDCS full off abruptly beyond the interval [$-50$, $+50$] and figure \ref{fig9} shows clearly that beyond $\pm 50$, the sum-rule is clearly checked.
\begin{figure}[!t]
 \centering
\includegraphics[angle=0,width=5 in,height=5 in]{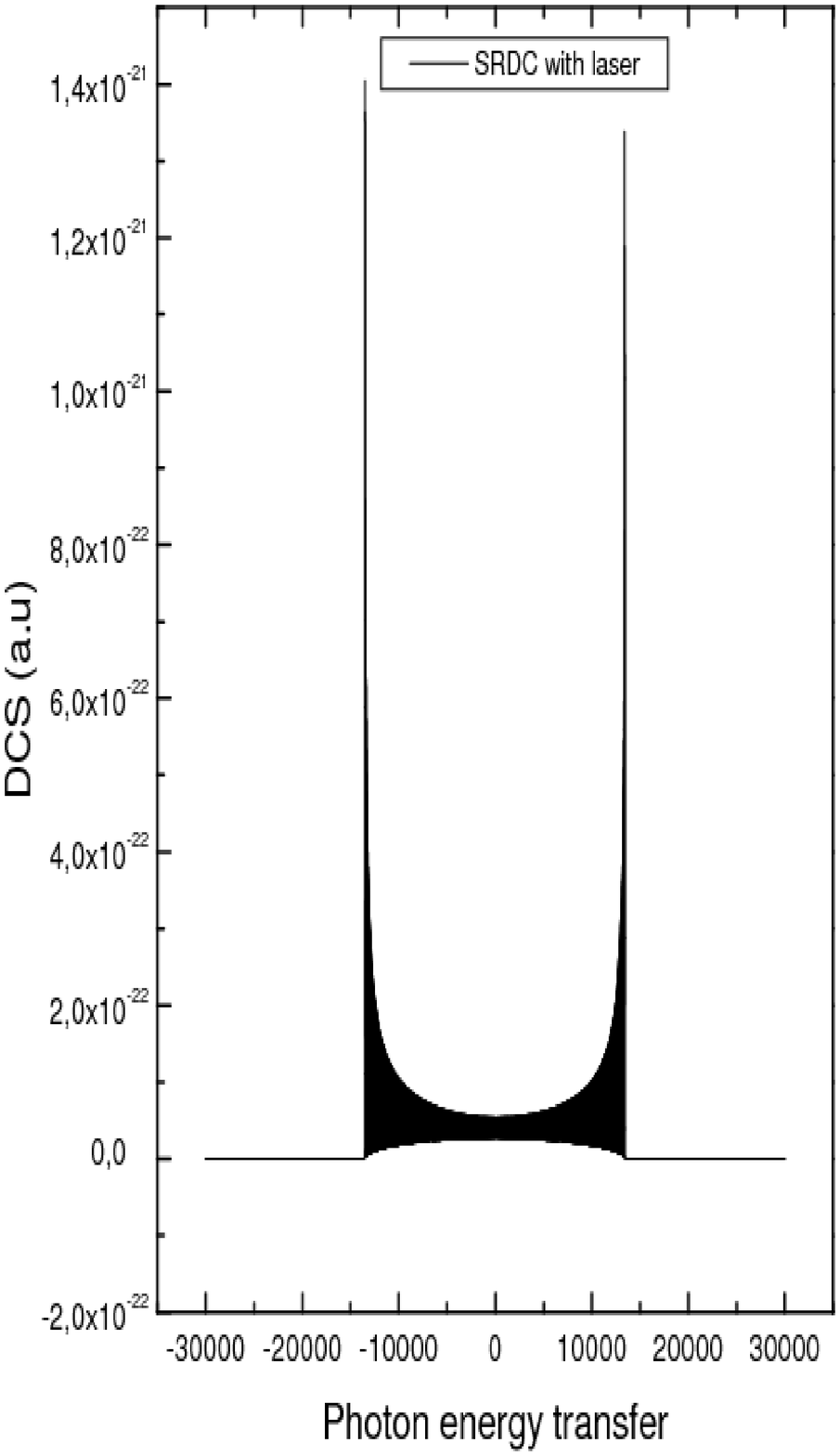}
\caption{\label{fig10} Envelope of the SRDCS in the relativistic regime ($\gamma=2.0$ and $\mathcal{E}=0.05\;a.u$) for a number of net photons exchanged $\pm 30000 $.}
\end{figure}
Figure \ref{fig10} show clearly that for the relativistic regime ($\gamma=2.0$ and $\mathcal{E}=0.05\;a.u$) and for the geometry ($\theta_i=45^{\circ}$, $\phi_i=0^{\circ}$, $\phi_f=45^{\circ}$ and $\theta_f=45^{\circ}$), the value of the cutoffs have been changed to $s=-13986$ for the negative part of the envelope and $s=+13912$ for the positive part. In this regime, when the number of photons exchanged increases, the convergence of the two SRDCSs (with and witout laser)   will be reached at approximatively $\pm 13945$, but as our computational capacity is limited, it is not possible, actually, to give the figure which illustrates such a situation.

\section{Conclusion}
This paper display the results of a semirelativistic
excitation of atomic hydrogen by electronic impact in the presence
and absence of the laser field. The
simple semirelativistic Darwin wave function that allows to obtain
analytical results in an exact and closed form within the framework
of the first Born approximation is used.  Our results have been compared
with previous nonrelativistic results, revealing that the
agreement between the different theoretical approaches is good in
the nonrelativistic regime. The
nonrelativistic treatment has been shown to be no longer reliable for high energies. It has been also found that a simple formal analogy links the
analytical expressions of the unpolarized differential cross section
in the absence of the laser field and the laser-assisted unpolarized differential cross section.


\begin{thebibliography}{99}
\bibitem{1} A. Weingartshofer, J. K. Holmes, J. Sabbagh and S. I. Chu, J. Phys. B \textbf{16}, 1805
(1983). See also, B. Wallbank, J. K. Holmes, J. Phys. B \textbf{27}, 1221 (1994).
\bibitem{2} M. A. Khaboo, D. Roundy and F. Rugamas, Phys. Rev. A \textbf{54}, 4004 (1996).
\bibitem{3} S. Luan, R. Hippler and H. O. Lutz, J. Phys. B \textbf{24}, 3241(1991).
\bibitem{4} N. J. Mason and W. R. Newell, J. Phys. B \textbf{22}, 777 (1989).
\bibitem{5} B. Wallbank, J. K. Holmes and A. Weingartshofer, Phys. Rev. A \textbf{40}, 5461, (1989);
J.Phys. B 23, 2997 (1990).
\bibitem{6} N. K. Rahman and F. H. M. Faisal, J. Phys. B \textbf{11}, 2003 (1978).
\bibitem{7} S. Jetzke, F. H. M. Faisal, R. Hippler and O. H. Lutz, Z. Phys. A 315, 271, (1984).
\bibitem{8} S. Jetzke, J. Broad and A. Maquet, J. Phys. B \textbf{20}, 2887 (1987).
\bibitem{9} R. S. Pundir and K. C. Mathur, Z. Phys. D \textbf{1}, 385 (1986).
\bibitem{10} F. W. Byron Jr, P. Francken and C. J. Joachain, J. Phys. B \textbf{20} 5487 (1987).
\bibitem{11} F. W. Byron Jr and C. J. Joachain, Phys. Rev. A 35 1590 (1987).
\bibitem{12} P. Francken, Y. Attaourti and C. J. Joachain, Phys. Rev. A \textbf{38}, 1785 (1988).
\bibitem{13} D. M. Volkov, Z. Phys 94, 250 (1935).
\bibitem{14} Y. Attaourti, B. Manaut and A. Makhoute, Phys. Rev. A \textbf{69}, 063407 (2004).
\bibitem{15} Alfred Maquet, Richard Taïeb and Valérie Véniard, Springer Series in Optical Sciences, Volume 134, 477-496, (2008).
\bibitem{16} R. Kisielius, K.A. Berrington and P.H. Norrington  J. Phys. B, \textbf{28}, 2459-2471, (1995).
\bibitem{17} J. Eichler and W.E. Meyerhof, \textit{Relativistic Atomic Collisions}, Academic Press,
(1995).
\bibitem{18} F.W. Jr Byron and C.J.\ Joachain, Phys. Rep. \textbf{179}, 211, (1989).
\bibitem{19} Gradstein, L S., Rizik, I.M. : Tables of Integrals, Sutures, Sets and Their Products.
Moscow: Nauka. (1971).
\bibitem{20} A. G. Grozin, Using REDUCE in High Energy Physics (Cambridge University, Cambridge, England,
1997).
\bibitem{20'} Y. Attaourti and B. Manaut Phys. Rev. A, \textbf{68}, 067401 (2003)
\bibitem{21} J. F. Williams, J. Phys. B \textbf{14}, 1197 (1981).
\bibitem{22} Bunkin F V and Fedorov M V Sov. Phys., JETP \textbf{22}, 284, (1966) ; Kroll N M and Watson K N Phys. Rev. A, \textbf{8}, 804, (1973).
\end{thebibliography}
\end{document}